\documentclass[12pt, draftclsnofoot, onecolumn]{IEEEtran}
	\usepackage{pifont,bm,multicol,amsfonts,amsmath,color,amssymb,graphicx, epsfig,cite,psfrag,subfigure,algorithm,enumerate,stfloats,algorithm,algorithmic,epstopdf,balance}
\usepackage{tabularx}
	\usepackage{pifont,bm,multicol,amsfonts,amsmath,color,amssymb,graphicx, epsfig,cite,psfrag,subfigure,algorithm,enumerate,stfloats,algorithm,algorithmic,epstopdf,balance}

\begin{document}
\title{Deep Learning based Channel Extrapolation for Large-Scale Antenna Systems: Opportunities, Challenges and Solutions} 	
\author{Shun Zhang, {\emph{Senior Member, IEEE,}} Yushan Liu, Feifei Gao, {\emph{Fellow, IEEE,}} Chengwen Xing, {\emph{Member, IEEE,}} Jianping An, {\emph{Member, IEEE,}} and Octavia A. Dobre, {\emph{Fellow, IEEE}}

%\thanks{S. Zhang and Y. Liu are with the State Key Laboratory of Integrated Services Networks, Xidian University, Xi'an 710071, P. R. China (Email: zhangshunsdu@xidian.edu.cn; ysliu\_97@stu.xidian.edu.cn).}
%
%\thanks{F. Gao is with Department of Automation, Tsinghua University, State Key Lab of Intelligent Technologies and Systems, Tsinghua University, State Key for Information Science and Technology (TNList) Beijing 100084, P. R. China (Email: feifeigao@ieee.org).}
%
%\thanks{C. Xing and J. An are with the School of Information and Electronics, Beijing Institute of Technology, Beijing 100081, China (e-mail: chengwenxing@ieee.org; an@bit.edu.cn).}
%
%\thanks{Geoffrey Ye Li is with the School of Electrical and Computer Engineering, Georgia Institute of Technology, Atlanta, GA 30332-0250 USA (e-mail: liye@ece.gatech.edu).}
%}

\thanks{S. Zhang and Y. Liu are with the State Key Laboratory of Integrated Services Networks, Xidian University, Xi'an 710071, P. R. China (Email: zhangshunsdu@xidian.edu.cn; ysliu\_97@stu.xidian.edu.cn). F. Gao is with Department of Automation, Tsinghua University, State Key Lab of Intelligent Technologies and Systems, Tsinghua University, State Key for Information Science and Technology (TNList) Beijing 100084, P. R. China (Email: feifeigao@ieee.org). C. Xing and J. An are with the School of Information and Electronics, Beijing Institute of Technology, Beijing 100081, China (e-mail: chengwenxing@ieee.org; an@bit.edu.cn). O. A. Dobre is with Faculty of Engineering and Applied Science, Memorial University, St. John's NL AIC-5S7, Canada (e-mail: odobre@mun.ca).}
}

\maketitle
\vspace{-16mm}
\begin{abstract}
With the depletion of spectrum, wireless communication systems turn to exploit large antenna arrays to achieve the degree of freedom in space domain, such as millimeter wave massive multi-input multi-output (MIMO), reconfigurable intelligent surface assisted communications and cell-free massive MIMO.
In these systems, how to acquire accurate channel state information (CSI) is difficult and becomes a bottleneck of the communication links.
In this article, we introduce the concept of channel extrapolation that relies on a small portion of channel parameters to infer the remaining channel parameters.
Since the substance of channel extrapolation is a mapping from one  parameter subspace   to another, we can resort to deep learning (DL), a powerful learning architecture, to approximate such mapping function.
Specifically, we first analyze the requirements, conditions and challenges for channel extrapolation.
Then, we present three typical extrapolations over the antenna dimension, the frequency dimension, and the physical terminal, respectively. We also illustrate their respective principles, design challenges and DL strategies.
It will be seen that channel extrapolation could greatly reduce the transmission overhead and subsequently enhance the performance gains compared with the traditional strategies. In the end, we provide several potential research directions on channel extrapolation for future  intelligent communications systems.

\end{abstract}

\section{Introduction}

With the rapid development of wireless technologies, we are entering the era of the fifth generation (5G) wireless communications and are heading towards the sixth generation (6G). Massive multi-input multi-output (MIMO) will continue to serve as one of the key technologies for 6G, as it can provide much more degrees of freedom than the conventional MIMO.
In particular, massive MIMO combined with millimeter wave (mmWave) communications can achieve orders of magnitude enhancement in system throughput. For the cell-free massive MIMO, the distributed systems can potentially reduce the inter-cell interference through coherent cooperation between base stations (BSs) and provide higher coverage than co-located massive MIMO.
Besides, the reconfigurable intelligent surface (RIS) with massive number of passive antennas can effectively perform the desired beamforming and reconstruct the radio scattering environment into an intelligent environment.
Unfortunately, as the number of antennas increases, all these promising wireless technologies require a large amount of training overhead in order to achieve accurate channel state information (CSI).

Traditionally, the channel reciprocity is utilized in time division duplex (TDD) massive MIMO systems to reduce the cost of downlink channel estimation as long as the uplink CSI is obtained.
For frequency division duplex (FDD) massive MIMO, the compressive algorithms that exploit the sparsity of channel in angle domain have been used to estimate CSI.
However, the performance of compressive sensing is restricted with the linear sparsity assumption that does not accurately match  the complex nonlinear sparse characteristics of the environment. Recently, Alkhateeb {\it et al.} revealed the deterministic relationship among different channels at different antennas and frequency bands, and introduced a new concept, namely the channel mapping in space and frequency domain \cite{channel_mapping}. With a similar principle, Esswie {\it et al.} proposed a spatial channel estimation scheme to reconstruct the downlink channel from uplink channel measurements for FDD MIMO systems \cite{ante_extra}. In \cite{cov_esti_sub6_mmWave}, Ali {\it et al.} constructed the channel covariance of the mmWave link from the spatial characteristics of the sub-6 GHz band.

In fact, the essential principle of deterministic mapping among different channels is that users at different spaces or different frequencies experience the same electromagnetic environment. Utilizing such deterministic mapping, the CSI at one space/frequency point can be used to predict the CSI at another space/frequency point, which is referred to as $\mathrm{\it channel~extrapolation}$.
Clearly, channel extrapolation can be excitingly useful to reduce the training cost over massive MIMO related systems.
%Although researchers may have been aware of the principle of channel extrapolation, mathematically describing the deterministic mapping among channels at different space/freqeuncy points has  never been achieved in existing  literature.
Inspired by the universal approximation capability of deep learning (DL) \cite{JinShi_DL,CaijunZhong_DL,Macro_DL}, it is possible to use DL to characterize the mapping among channels at different space/frequency locations.

In this article, we introduce the mechanism for channel extrapolation and analyze its major challenges over three scenarios: antenna domain extrapolation, frequency domain extrapolation and physical terminal extrapolation.
Specifically, for the antenna extrapolation, we propose to use the channel of a few uplink antennas to reconstruct the full downlink channel for an FDD system, and present the corresponding DL-based solution.
For the frequency extrapolation, we consider two cases: the channel mapping between two subcarrier sets within one frequency band and the channel mapping across different frequency bands.
For the terminal extrapolation, we consider the channel mapping among different distributed users, where the sensors are applied to provide additional information about the environment and DL is used to achieve the transfer between neural networks (NNs) of distinct users. Finally, we discuss several potential research directions on channel extrapolation for future intelligent communications.

\section{Problem Formulation}
\subsection{The Concept of Channel Extrapolation}
\begin{figure}[!t]
	\centering
	\includegraphics[width=6in]{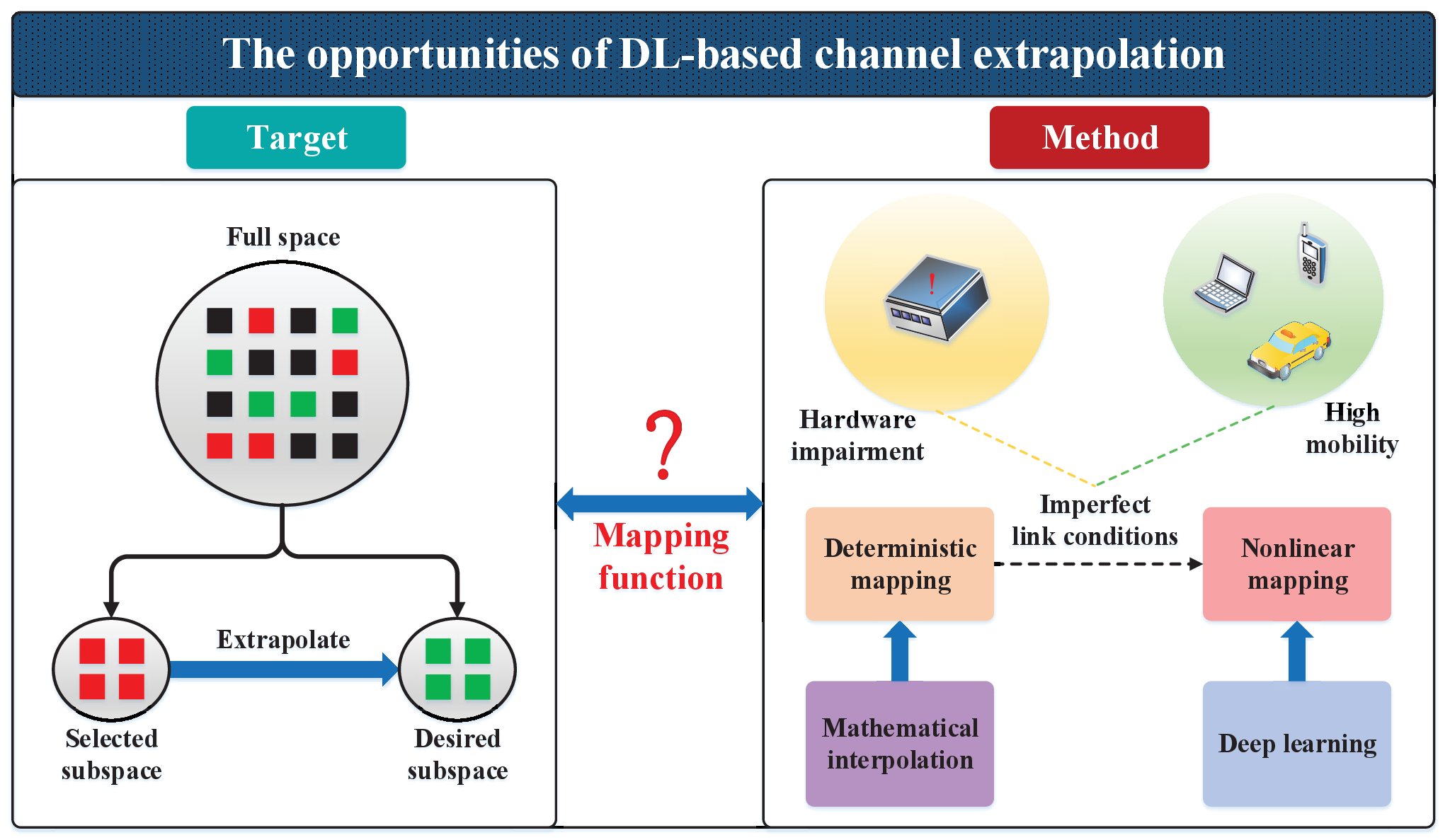}
	\caption{The opportunities of DL-based channel extrapolation.}
	\label{fig:channel_extrapolation}
\end{figure}

From the propagation characteristics of the electromagnetic wave, we know that when the user switches the frequency or moves in a short time, its  surrounding environment would not change much, which makes the channels at different spaces or different frequencies to have certain mapping relationship. Many efforts have been made to explore such mapping, and some effective mapping models, e.g., the one between the uplink and downlink channels over the massive MIMO systems, have been established.
Nevertheless, channel mappings, like the one between the sub-6 and mmWave bands or the one among different distributed antennas, are difficult to describe mathematically.
%\textcolor[rgb]{1.00,0.07,0.00}{Even though the accurate mapping model that can be built theoretically can be solved by utilizing the signal reconstruction, the compressed sensing and the mathematical interpolation theories, under the complicated scenarios, such as hardware impairment and high mobility, the exact mapping model can not be easily achieved, and the model based schemes can not work effectively. û¿´¶®}
In \cite{channel_mapping}, the authors proved that the large-scale antenna arrays are able to assure the \emph{bijectiveness} condition, which makes the channel mapping along the frequency and space dimensions unique. Then, the authors utilized the DL approach to successively approximate such channel mapping.
%Note that using \textcolor[rgb]{0.00,0.07,1.00}{one channel subspace to infer the CSI in another channel subspace} is named as $\mathrm{\it %channel~extrapolation}$ and utilizing the DL to realize the channel extrapolation is named as based $\mathrm{\it DL}$-$\mathrm{\it based~channel~extrapolation}$.
The key idea of DL-based channel extrapolation is illustrated in Fig.~\ref{fig:channel_extrapolation}.

\subsection{Challenges of DL-based Channel Extrapolation}
In fact, channel extrapolation can be treated as a mapping from one parameter subspace  to another, which depends on three aspects: the acquisition of the original subspace information, the choice of the original subspace, and the mapping scheme from the original subspace to the targeted one.
Notice that the task of ``the original subspace acquisition'' is to estimate the sub-sampled channel of small size and can be easily implemented through the conventional Bayesian linear estimator, message passing algorithms, or DL-based algorithms.
Therefore, in the following, we will focus on the other two challenges:

\begin{enumerate}
\item {\it{Subspace selection}}: For the given full space, the raw input information is closely related with the selection of subspace. Different selections would correspond to distinguished information compression rates and  $\mathrm{\it selection~patterns}$, where the selected space is marked as `1' while the others are marked as `0'. With the power,  hardware, or performance constraints, it is possible to optimize the selection pattern and to provide a good start point for a specific extrapolation task. The sparser the selection pattern is, the less the training costs and hardware power consumption are, but the poorer the performance of the channel extrapolation will be.
    Thus, with the DL-based channel extrapolation, it is important to determine the selection pattern before starting the transmission.
    However, the core operation of DL is the gradient descent algorithm that requires the mapping function to be continuous and differentiable. Hence, the optimization of the subspace selection is challenging for DL based channel extrapolation.

\item {\it{Mapping scheme}}: The DL-based channel extrapolation is similar to the super-resolution in the field of image processing, in which it is important to properly exploit the correlation between data elements for information completion.
    In order to improve the performance of the NN, we can increase the number of data layers or modify the NN structure.
    However, more layers result in heavier calculation, and when the number of layers reaches a certain number, the degree of improvement becomes smaller. Sometimes, excessively deepening the NN will even cause the gradient explosion and disappearance.
    Thus, we aim to constructing a robust but simple NN structure to achieve better extrapolation performance and faster convergence, which is another challenge of the channel extrapolation.
\end{enumerate}

In the following, we present the channel extrapolation over the antenna dimension, the frequency dimension and the physical terminal separately, and show the effectiveness of each type of extrapolation.

\section{Channel Extrapolation over Antenna Dimension}
\begin{figure}[!t]
	\centering
	\includegraphics[width=6in]{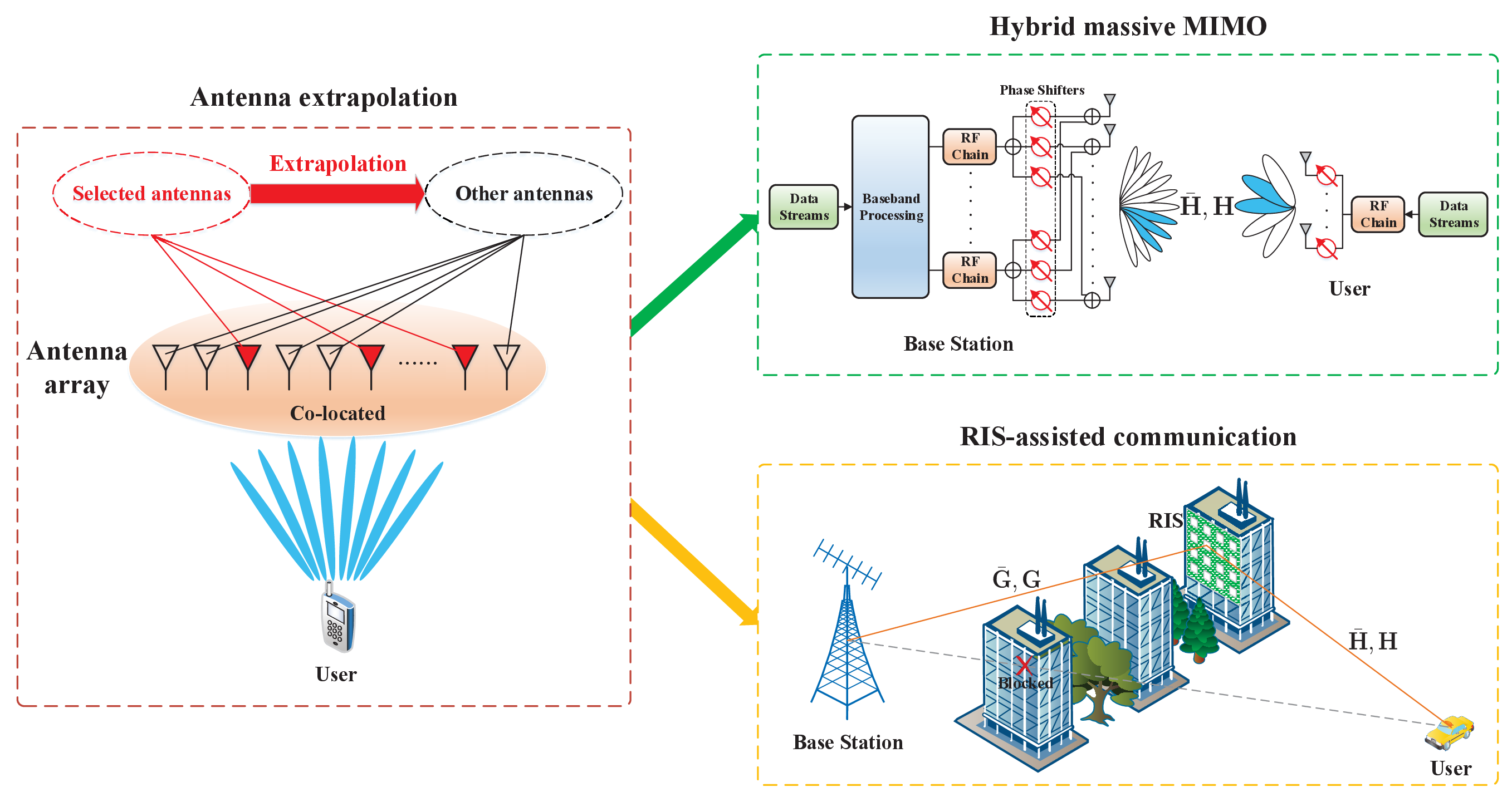}
	\caption{The channel extrapolation over antenna dimension for large-scale antenna systems.}
	\label{fig:antenna_extrapolation}
\end{figure}

A practical way to implement large antenna arrays, e.g., massive MIMO, is to use the hybrid analog and digital architecture, where a small number of radio frequency (RF) chains are connected to massive antennas through the fully connection or subarray connection structure. In this case, the limited number of RF chains have to connect the active antennas in turn for the acquisition of downlink CSI, which consumes significant time resources, especially when physically switching the antennas costs non-ignorable time.
Besides, over the  RIS communication network, the channel size is in scale with the number of RIS elements, which is usually very large to achieve good electromagnetic reconfiguring performance. Thus, the channel estimation in RIS-aided network also requires significant time resources.
With the new idea of channel extrapolation, we can explore the mapping relation between the partial antennas and the full one in an offline manner, and accurately infer the CSI of full antennas from the partial antennas during the online transmission. This type of channel extrapolation architecture is specifically referred to as $\mathrm{\it antenna~extrapolation}$ and is illustrated  in Fig. \ref{fig:antenna_extrapolation}.

Recently, there are some preliminary results on implementing the antenna extrapolation.
In \cite{yindiyang}, the authors examined the antenna extrapolation for a massive MIMO system based on the deep NN (DNN), which can capture the inherent relationship between the uplink and downlink channel data sets and extrapolate the downlink channels from a subset of the uplink CSI.
In \cite{shunbo zhang}, Zhang {\it et al.} realized the antenna extrapolation through a convolutional NN (CNN) in a RIS-aided communication system. Moreover, the authors proposed an antenna selection network that utilizes the probabilistic sampling theory to select the optimal locations of those active antennas.
In \cite{ODE_CNN}, we modify the NN structure by ordinary differential equation (ODE) which can describe the latent relations among different data layers and improve the performance gains of the antenna extrapolation.

\begin{figure}[!t]
	\centering
	\includegraphics[width=6in]{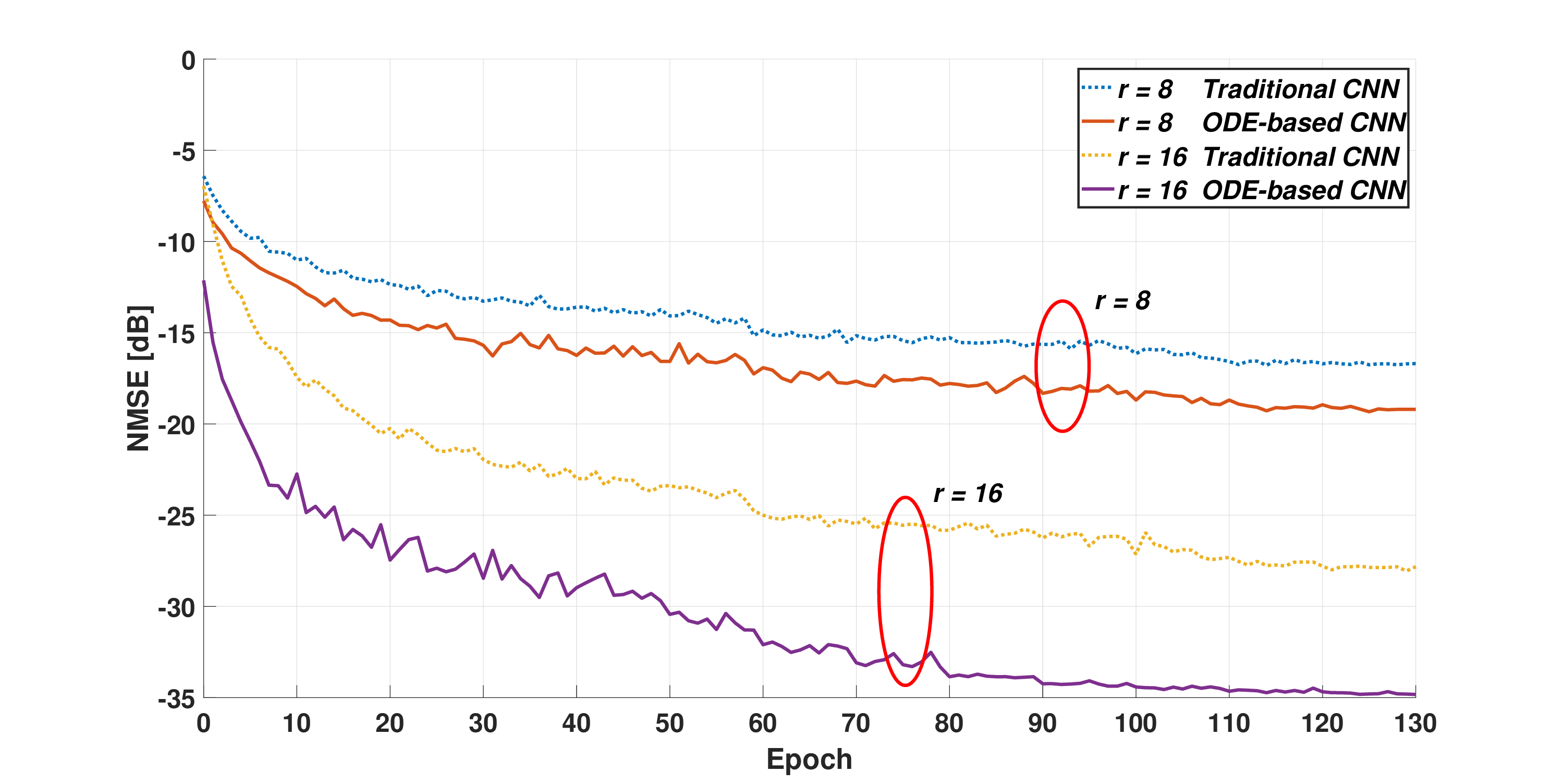}
	\caption{{The NMSE comparison of traditional CNN and ODE-based CNN against epochs for antenna extrapolation in a RIS-assisted communication system. Note that $r$ denotes the number of active antennas at RIS.}}
	\label{fig:antenna_plot}
\end{figure}

In Fig. \ref{fig:antenna_plot}, we provide evaluation results of the antenna extrapolation for the traditional CNN structure and the ODE-based CNN structure in a RIS-aided system.
The parameter settings are: a $4\times 4$ uniform planar array (UPA) at BS, $8\times 8$ UPA at RIS and single antenna at the user, where the antenna spacing is half wavelength.
The carrier frequency is $2.4$ GHz, the system bandwidth is $20$ MHz, and the number of subcarriers is $64$. Moreover, the number of active antennas at RIS is separately taken as $8$ and $16$. The distribution of training and test users in the DeepMIMO dataset is the same as in \cite{yindiyang}.
It can be seen from Fig. \ref{fig:antenna_plot} that with the increase of iteration time, the normalized mean square error (NMSE) curves decay fast first, and then slow down after $110$ epochs. As the number of active antennas increases, the NMSEs of the two structures decrease and the performance gap between the two structures becomes larger. Hence, for a given number of active antennas, the ODE-based CNN outperforms the traditional CNN and trains faster than the traditional CNN.

In fact, the DL-based algorithms should adapt to the environmental changes and customize the antenna extrapolation schemes according to the environmental information.
Hence, once the environment changes, say when the users are moving, the corresponding NNs should be re-trained.
Thus, how to achieve a good balance between the performance gains and the cost caused from retraining the DNN is another challenge for the antenna extrapolation.
A possible solution could be the transfer and meta learning that has the ability to adapt to the new environment quickly with a small amount of new training samples \cite{transfer_learning}.

\section{Channel Extrapolation over Frequency}
\begin{figure}[!t]
	\centering
	\includegraphics[width=6in]{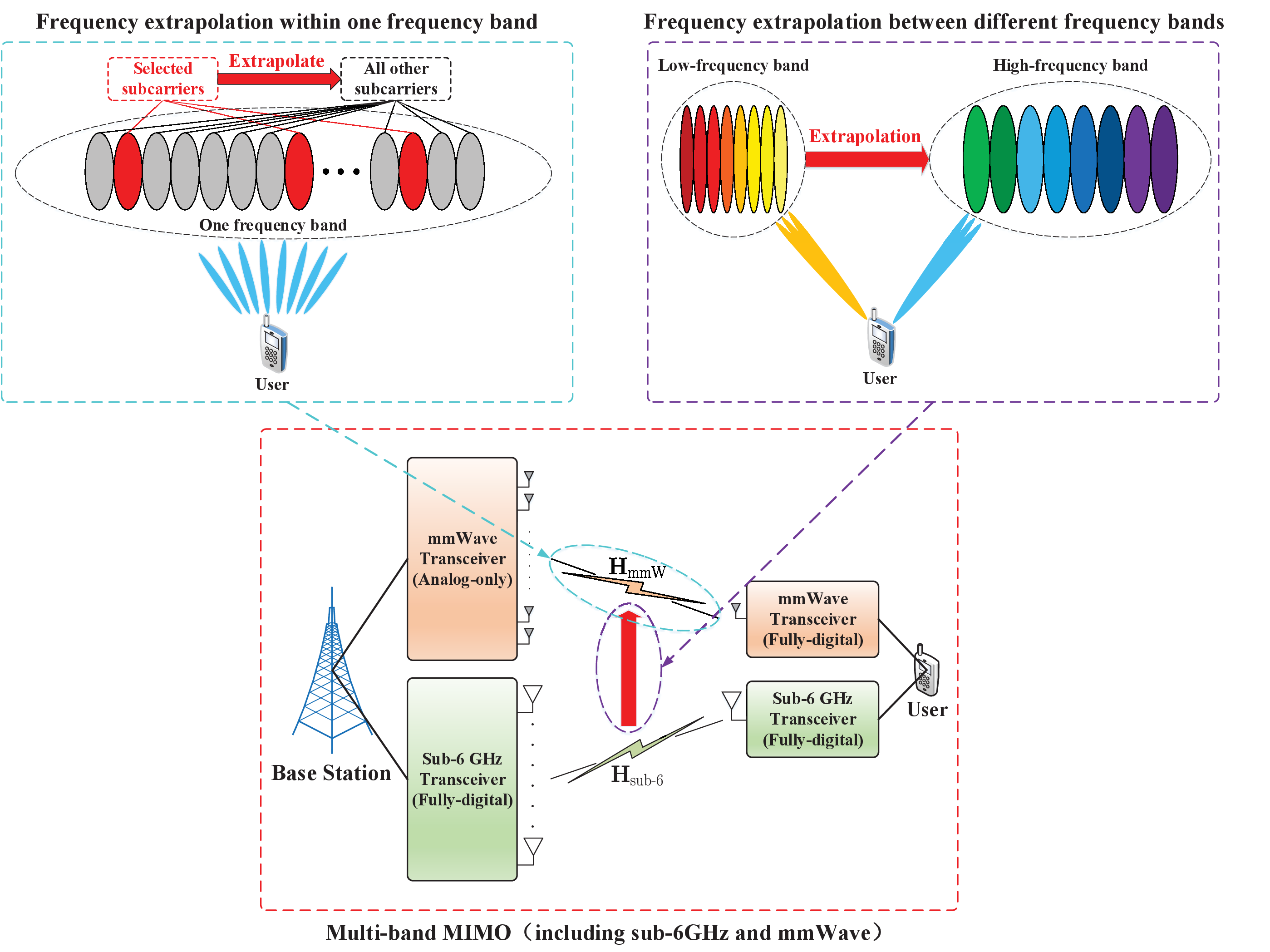}
	\caption{The channel extrapolation over frequency for multi-band MIMO systems, including sub-6 GHz and mmWave.}
	\label{fig:frequency_extrapolation}
\end{figure}

Along the frequency dimension, the channel extrapolation can use one set of subcarriers to infer another set of subcarriers, which is named $\mathrm{\it frequency~extrapolation}$. We present two typical applications for the frequency extrapolation in Fig. \ref{fig:frequency_extrapolation}. One is to implement the channel extrapolation between two subcarrier sets within a given frequency band. The other is to extrapolate channels among different frequency bands, being suitable for the multi-band systems, say FDD massive MIMO system. As seen later, the frequency extrapolation principle may even be feasible when the gap between different frequency bands is large.

\subsection{Frequency Extrapolation Within A Frequency Band}
The orthogonal frequency division multiplexing (OFDM) technology with a large amount of subcarriers is usually adopted to capture the potential efficiency of large bandwidth and deal with the frequency selective fading.
Since the mathematical channel modeling among different subcarriers is available, one can sparsely place pilots at partial subcarriers, sound the incomplete channels and reconstruct the entire CSI over the full band via a signal processing approach.
Although the performance of the signal processing approach is very good in a stable environment, its effectiveness degrades in more complicated scenarios. For example, in high mobility scenario, OFDM may encounter significant inter-carrier interference due to the Doppler spread. Moreover, when  the length of the predefined cyclic prefix (CP) is smaller than that of the channel's delay spread, the inter-symbol interference appears and destroys the signal modeling in the frequency domain. In addition, the hardware impairments, such as the I/Q imbalance, phase noise and nonlinearity of the power amplifier, would cause nonlinear distortion in received signals.
%Under the above complicated scenarios, the mapping among different subcarriers channels becomes intricate, which may make the uniform pilot pattern and the mathematical interpolation not the optimal scheme anymore.

A possible solution is to utilizing the universal approximation capability of  DL to characterize the channel modeling in these complicated scenario and continue to reconstruct the channels of all subcarriers from a small subset of subcarriers. Another key problem is that the traditionally adopted uniform pilot pattern cannot guarantee the optimal performance anymore and the design of the pilot pattern should be carefully addressed.

\begin{figure}[!t]
	\centering
	\includegraphics[width=6in]{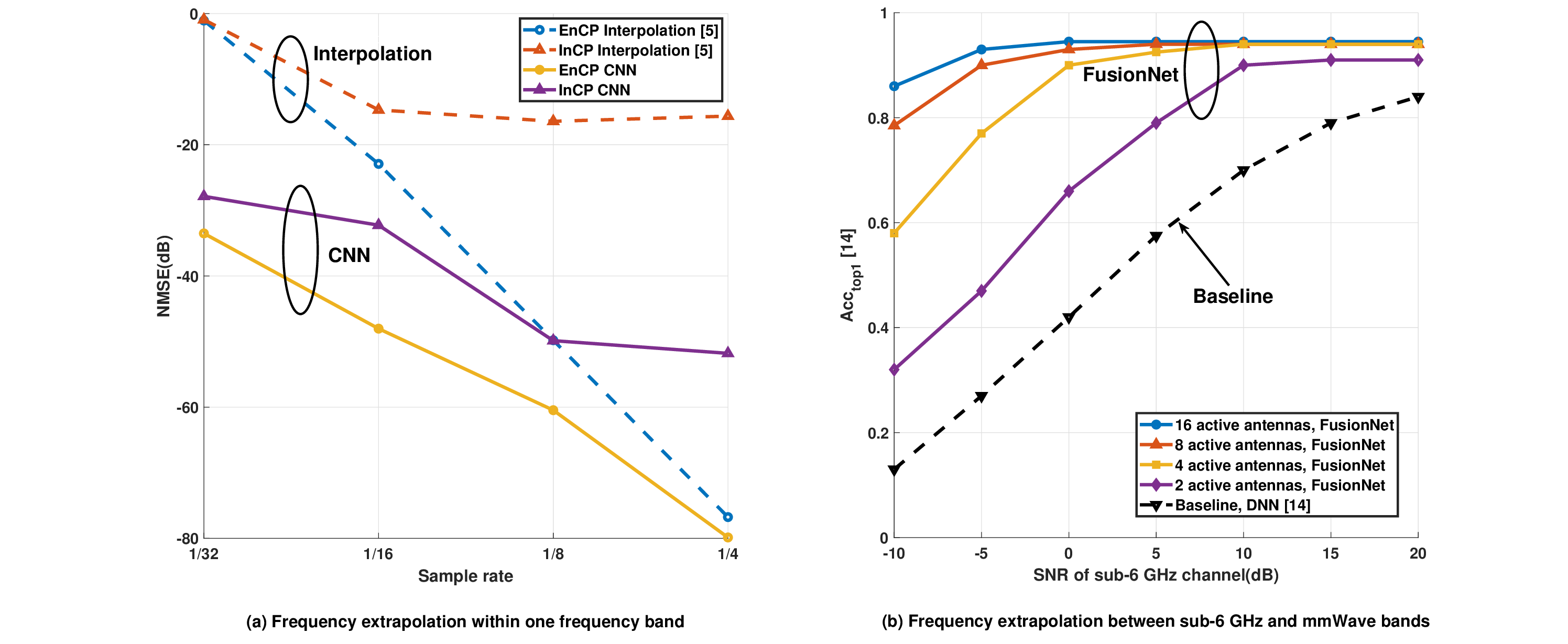}
	\caption{The performance analysis of two frequency extrapolations in a multi-band system (including sub-6 GHz and mmWave bands).}
	\label{fig:frequency_plot}
\end{figure}

In Fig. \ref{fig:frequency_plot} (a), we offer the evaluation results of the frequency extrapolation in mmWave band based on the CNN and conventional interpolation scheme \cite{CaijunZhong_DL}. The antenna configuration is the same as that in Fig. \ref{fig:antenna_plot}. The carrier frequency is $60$ GHz, and the subcarrier spacing is $60$ kHz. The number of subcarriers is $1024$, and the number of selected subcarriers for partial channel estimation is $256$. The curves labeled by `InCP' correspond to the case of insufficient CP, while the ones marked by `EnCP' represent the case of enough CP.
It can be seen from the Fig. \ref{fig:frequency_plot}(a) that the NMSE curves decay with the increase of the sample rate on the subcarrier. Meanwhile, the NMEs of CNN are always better than those of conventional interpolation, especially in the case of insufficient CP. Moreover, in the case of enough CP, the gap between CNN and conventional interpolation becomes larger as the sample rate decreases.

\subsection{Frequency Extrapolation Between Different Frequency Bands}
DL-based frequency extrapolation can also be used to infer the downlink channel from the uplink one in FDD massive MIMO, which solves the problem resulted from the fact that the channel is not reciprocal in the two different frequency bands. In fact, the technology of multiple bands transmission is a new trend, especially that recent communications systems are designed to work at both sub-6 GHz and mmWave bands, simultaneously  \cite{cov_esti_sub6_mmWave,sup_sub6_mmWave,beam_extra_DL}. In the dual band system, the mmWave spectrum would provide a high speed link and offer gigabit-per-second data rates, but it would also face huge training overhead and high sensitivity to blockages. Since the mmWave and sub-6 GHz links experience the same scattering environment, it is possible to build a deterministic relationship between their channels and extrapolate the CSI of mmWave band from that of the sub-6 GHz band.

In \cite{sup_sub6_mmWave}, Li {\it et al.} used the sub-6 GHz spatial information to help beam selection in {a mmWave system} by exploiting the feature that the support set of {the} mmWave channels is a subset of that for the sub-6 GHz channels with the same grid quantization.
In \cite{beam_extra_DL}, Alrabeiah {\it et al.} performed beam prediction at mmWave band from the CSI at sub-6 GHz band with {the aid of a DNN}. Moreover, {the authors} incorporated the materials' dielectric coefficients at the sub-6 GHz and mmWave bands, which {is critical information for} the accurate mapping between {the} sub-6 GHz and mmWave signals.

Although the effectiveness of the above work has been verified, the frequency extrapolation from sub-6 GHz  to  mmWave is practically inaccurate. This is mainly because the mmWave signal propagation is highly sensitive to blockages, and the electromagnetic microwave impinging on the arrays of two links have different angular spread (AS).
Therefore, the mmWave channel is a sub-category of the sub-6 GHz channel \cite{relation_sub6mmWave}, and the CSI of mmWave band could seriously deviate from that in sub-6 GHz. In order to calibrate the CSI deviation, we design a simple but tricky dual-input {NN, referred to as} FusionNet, to merge the features of sub-6 GHz channels and a few inherent pilots at mmWave band to improve the beam prediction \cite{beam_pilot_DL}. It is worth mentioning that {the number of mmWave pilots is} not enough to meet the channel estimation {requirements; however, it} can help refine the mmWave beam direction on top of the CSI information from {the} sub-6GHz band.
Moreover, the balance between the auxiliary pilot overhead and the extrapolation performance can be further optimized, and the pilot pattern along the frequency dimension can be improved through the probabilistic sampling theory.

In Fig. \ref{fig:frequency_plot} (b) we display the evaluation results of {the} mmWave beam prediction based on DNN and FusionNet, respectively.
The parameter configuration and the definition of top-1 accuracy $Acc_{top1}$ are {the same as in} \cite{beam_pilot_DL}.
The `baseline' curve is the performance of the DNN, and the `active antennas' represent the mmWave antennas that participate in the mmWave channel estimation.
The number of active antennas in {the} mmWave system is taken as $2,4,8,16$ respectively, and the mmWave pilot {signal-to-noise ratio (SNR)} is $20$ dB.
It can be seen from the Fig. \ref{fig:frequency_plot}(b) that the prediction accuracy of the FusionNet with any number of mmWave active antennas is always better than {that of} the baseline method, especially {at low SNR for} the sub-6 GHz channel estimation.
Moreover, as the number of active mmWave antennas increases, the beam prediction accuracy of  FusionNet improves significantly but slows down after the number of active mmWave antennas exceeds $8$.

\section{Channel Extrapolation over Physical Terminal}

\begin{figure}[!t]
	\centering
	\includegraphics[width=6in]{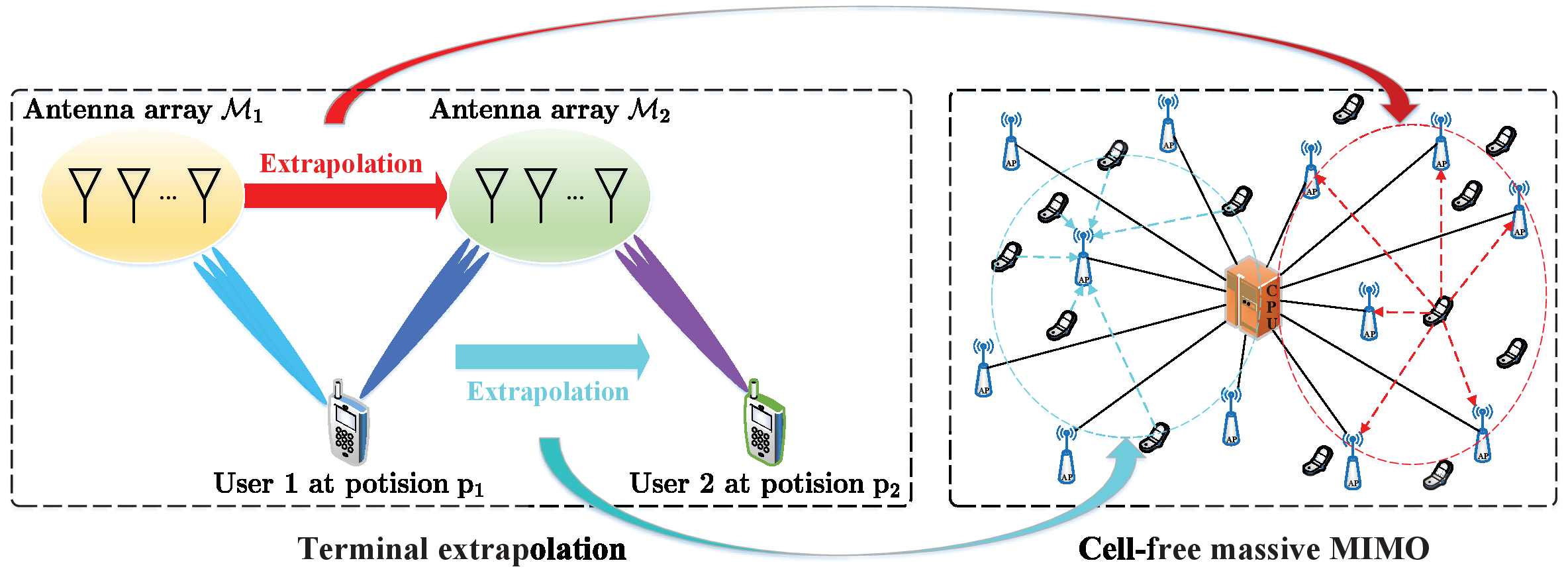}
	\caption{The channel extrapolation over the physical terminal for a cell-free massive MIMO system. Note that the terminal can be antenna array and user.}
	\label{fig:space_extrapolation}
\end{figure}

Inspired by the channel-to-channel mapping, the authors of \cite{channel_mapping} verified that the channel extrapolation among different terminals at any position is possible, and then applied this mapping concept to the distributed (cell-free) massive MIMO. With the key {idea of} \cite{channel_mapping}, one can {employ} a subset of terminals to infer the information {for the other} terminals, which is named $\mathrm{\it terminal~extrapolation}${; this is illustrated} in Fig. \ref{fig:space_extrapolation}.
If the terminals are close or are in a similar environment, then the terminal extrapolation is feasible and {provides} enough accuracy. However, with the increase of the distance between terminals, the number of common scattering objects becomes smaller and the latent relation becomes weaker, which {deteriorates} the performance of the terminal extrapolation.
%However, the distance of distributed antennas in cell-free massive MIMO system or that of the different users over multi-user system is often relatively far.
%Then, the channel mapping function among different antenna sets or distinct users may be intricate and indistinct.
In addition, this mapping would be closely related with {the} link conditions, such as physical position of terminals and hardware states, which {causes} difficulties for terminal extrapolation.
Hence, unlike the antenna and frequency extrapolations, {the terminal extrapolation occurs among different parameter subspaces with large differences and it is difficult to realize satisfactory extrapolation performance only through NNs}. %One feasible strategy it to perform extrapolation among the terminals with similar electromagnetic scattering environment while implement the transfer among the neural networks in different regions.

Meanwhile, with the application of various types of sensors, {the communication systems can avail of diverse information}, such as users' positions and mobility states.
Hence, one can incorporate the information from sensors to enhance the performance of terminal extrapolation \cite{Sensory_yuwen}.
Explicitly, the terminals are equipped with sensors to detect the map of the static environment.
Then, {they can be divided into groups of terminals that share similar electromagnetic scattering environments}.
Similar to the frequency extrapolation, we can resort to the FusionNet to merge a few pilots and sensing information to achieve {a good extrapolation} among terminals in one group.
Nevertheless, the terminals' channels in different groups are less correlated, and {thus, these groups} cannot share the common extrapolation {NN}.
Separately designing and training the extrapolation {NNs} for different groups would {yield a large} computation burden.

Inspired by the humans' ability to transfer knowledge from previous experience, transfer learning {has become a promising technology in the field of machine learning to solve similar tasks with limited labeled data. This aims to improve the performance of target tasks by exploiting the knowledge from source tasks.}
Recently, Yang {\it et al.} utilized {a meta-learning} scheme and effectively {adapted} the CSI estimation {NN} to the new environment \cite{transfer_learning}, which {provides} a feasible way to apply  transfer learning or meta-learning to the terminal extrapolation. Given the constraints on the performance of the terminal extrapolation, one should also determine which group should be selected to infer the {NNs} of {the other} groups. Moreover, the size of the group should be carefully designed to balance the performance of terminal extrapolation and training cost.

\section{Future Research Directions}
Although channel extrapolation has been established in theory and some preliminary results have been presented, there are still many open problems that need to be investigated over the large-scale antenna systems. We highlight several potential research directions as follows:

\subsection{Joint Channel Extrapolation over Antenna, Frequency and Physical Terminal}
The previously introduced three different channel extrapolation can be designed in a joint manner. For example, one can apply the OFDM system with large bandwidth into RIS-assisted communications, i.e., the combination of antenna and frequency extrapolation. In such case, one needs to implement 2D sampling along the antenna-frequency domain, and then use the small 2D subspace to construct the full 2D space over the antenna and frequency dimensions.

\subsection{Optimization of NN and Resource Deployment}
The existing works mainly demonstrated the effectiveness of DL-based channel extrapolations, while further effort is needed to optimize the parameters of the NN, e.g., the number of layers, as well as to consider optimal resource deployment, e.g., location of auxiliary mmWave pilots over frequency extrapolation,  and distributed antennas/users scheduling over terminal extrapolation.

\subsection{Model-Driven Channel Extrapolation}

The previously discussed channel extrapolation mainly works in a data-driven manner. If the link conditions are good, the mapping function among different subspaces may be modeled mathematically, and a model-driven approach may be designed to enhance the performance of channel extrapolation. We may also design schemes that could intelligently switch between the data-driven and model-driven manners according to the complexity requirement and the dynamic environment.

\subsection{Vision-based Channel Extrapolation}
In general, the communication environment information, such as building location, obstacle shape and material, dynamic pedestrian, and vehicle distribution, would determine the propagation of the electromagnetic waves at any location and any frequency band. Hence, one may resort to sensors such as camera or radar to capture the 3D scene image of the environment and then design the corresponding DL algorithm to assist all three types of channel extrapolation.

\section{Conclusion}

In this article, we have presented the channel extrapolation concept and analyzed its three major challenges, i.e., the acquisition of the original subspace information, the selection of the original subspace and the mapping scheme from the original subspace to the targeted one. We divided the channel extrapolation into three typical types: antenna extrapolation, frequency extrapolation and terminal extrapolation.
For antenna extrapolation, we found that the ODE-based NN outperforms the traditional NN model.
For frequency extrapolation, when the gap between different frequency bands is large, we need a small number of pilots to refine the extrapolation.
For terminal extrapolation, due to subspaces with large difference, it is difficult to achieve a good performance only though NNs. Hence, the utilization of sensory data and transfer learning can improve the extrapolation performance. Finally, we have introduced several potential research directions on channel extrapolation.

In summary, channel extrapolation is a very promising and powerful tool to handle the complicated transmission scenarios when the mathematical modeling of the channel is inaccurate or even unavailable.
This is especially applicable to mmWave massive MIMO, RIS-assisted communications and cell-free massive MIMO, and deserves a full investigation and exploitation for such communications systems, which will incorporate intelligence in the future.

%\balance

%\begin{thebibliography}{1}
%
%	\bibitem{metis6wireless}
%	M.~METIS, ``wireless communications enablers for the twentytwenty information
%	society, eu 7th framework programme project,'' ICT-317669-METIS, Tech. Rep.
%	\end{thebibliography}

%\balance
%\bibliographystyle{IEEEtran}
%\bibliography{./bibtex/IEEEabrv,./bibtex/ref}

%\balance
%\bibliographystyle{IEEEtran}
%\bibliography{./bibtex/IEEEabrv,./bibtex/ref}
\end{document}